\documentclass[english,aps,superscriptaddress,twocolumn]{revtex4}
\usepackage[T1]{fontenc}
\usepackage[latin1]{inputenc}
\usepackage{graphicx}

\usepackage{babel}
\makeatother

\begin{document}
\title{Suppression of spin-state transition in epitaxially strained LaCoO$_{3}$}

\author{C. Pinta}
\affiliation{Forschungszentrum Karlsruhe, Institut f\"{u}r Festk\"{o}rperphysik, 76021 Karlsruhe, Germany}
\affiliation{Physikalisches Institut, Universit\"{a}t Karlsruhe, 76128 Karlsruhe, Germany}
\author{D. Fuchs}
\author{M. Merz}
\affiliation{Forschungszentrum Karlsruhe, Institut f\"{u}r Festk\"{o}rperphysik, 76021 Karlsruhe, Germany}
\author{M. Wissinger}
\author{E. Arac}
\affiliation{Forschungszentrum Karlsruhe, Institut f\"{u}r Festk\"{o}rperphysik, 76021 Karlsruhe, Germany}
\affiliation{Physikalisches Institut, Universit\"{a}t Karlsruhe, 76128 Karlsruhe, Germany}
\author{A. Samartsev}
\affiliation{Physikalisches Institut, Universit\"{a}t Karlsruhe, 76128 Karlsruhe, Germany}
\affiliation{Forschungszentrum Karlsruhe, Institut f\"{u}r Festk\"{o}rperphysik, 76021 Karlsruhe, Germany}
\author{P. Nagel}
\affiliation{Forschungszentrum Karlsruhe, Institut f\"{u}r Festk\"{o}rperphysik, 76021 Karlsruhe, Germany}
\author{S. Schuppler}
\affiliation{Forschungszentrum Karlsruhe, Institut f\"{u}r Festk\"{o}rperphysik, 76021 Karlsruhe, Germany}
\date{\today}

\begin{abstract}
Epitaxial thin films of LaCoO$_{3}$ (E-LCO) exhibit ferromagnetic order with a transition temperature
$T_{\rm{C}}$~=~85~K, while polycrystalline thin LaCoO$_{3}$ films (P-LCO) remain paramagnetic. The temperature-dependent
spin-state structure for both E-LCO and P-LCO was studied by x-ray absorption spectroscopy at the Co~$L_{2,3}$ and O~$K$
edges. Considerable spectral redistributions over temperature are observed for P-LCO\@. The spectra for E-LCO, on the
other hand, do not show any significant changes for temperatures between 30 K and 450 K at both edges, indicating that
the spin state remains constant and that the epitaxial strain inhibits any population of the low-spin ($S = 0$) state
with decreasing temperature. This observation identifies an important prerequisite for ferromagnetism in E-LCO thin
films.
\end{abstract}


\maketitle

\section{Introduction}

The great variety of mutually competing phases in the cobaltates, caused by the large number of interactions occurring
on similar energy scales (e.g., Hund\textquoteright{}s coupling, crystal field, double exchange, and correlation) has
been attracting intense interest for decades. Despite such efforts, many of their aspects continue to stir discussion
and to spawn further, more detailed studies (see for instance Ref.s
\citep{Saitoh1997,Yamaguchi1997,Imada1998,Asai1989,Fauth2001,Zobel2002,English2002,Ravindran2002}). To a certain degree, cobaltates may appear similar to the manganites $-$ concerning structural aspects, for instance, or the partial
occupation of the $3d$ shell $-$, yet the fact that two or even three different spin states are so closely spaced
in energy that they are almost degenerate in many situations gives them their own distinct flavor and causes intriguing
physical properties: Bulk LaCoO$_{3}$, for instance, is a nonmagnetic semiconductor at low temperature,
T~\ensuremath{\le}~35~K, with the Co$^{3+}$ ions in a low-spin (LS) configuration ($t_{2g}^{6}e_{g}^{0}$, $S$=0)\@. A
transition of all Co$^{3+}$ ions to a high-spin state (HS, $t{}_{2g}^{4}e_{g}^{2}$, $S$=2) occurs at about 500 K and is
concurrent to an insulator-metal transition. In the temperature range 35~K~<~T~<~100~K where the susceptibility has a
broad maximum no consensus seems to have been reached yet: many authors
\citep{Saitoh1997,Saitoh1997a,Stolen1997,Ravindran2002,Zobel2002,Maris2003} interpret their experimental results in
terms of an intermediate-spin state (IS, $t{}_{2g}^{5}e_{g}^{1}$, $S$=1) as suggested by an LDA+U calculation
\citep{Korotin1996} (using the local density approximation, LDA, with on-site Hubbard-like correlation effects added),
while other groups understand the spin-state transition in this region as a gradual population of higher spin states
starting from LS \citep{Zhuang1998,Noguchi2002,Ropka2003,Haverkort2006}\@. In Ref.\ \citep{Haverkort2006}, a
configuration-interaction CI cluster-model calculation including the full atomic multiplet theory and hybridization
effects with the O 2\emph{p} orbitals is used to explain their Co $L{}_{2,3}$ x-ray absorption and x-ray magnetic
dichroism results; there is also good agreement with susceptibility data. The authors do not find an IS contribution in
their calculations. Generally accepted is, however, that crossovers between spin states occur in LaCoO$_{3}$ upon
changing the temperature. They are caused by the delicate balance between the crystal-field splitting $\Delta_{CF}$ and
the exchange interaction $J_{ex}$ due to the Hund's rule coupling, inducing a redistribution of electrons on the
$t_{2g}$ and $e_{g}$ levels. Since $\Delta_{CF}$ is obviously very susceptible to the Co-O bond length, the balance
between $\Delta_{CF}$ and $J_{ex}$ can easily be tipped to either side by doping or pressure.

In a previous work \citep{Fuchs2007} it was reported that in contrast to bulk material, epitaxial thin films of
LaCoO$_{3}$ (E-LCO) become ferromagnetic below the Curie temperature $T_{\rm{C}}$~=~85~K when grown on
(LaAlO$_{3}$)$_{0.3}$(Sr$_{2}$AlTaO$_{6}$)$_{0.7}$ (LSAT) substrates, while polycrystalline thin films (P-LCO) grown
under similar conditions do not show any ferromagnetism down to T~\ensuremath{\approx}~5~K\@. The magnetic behavior of
the films has been explained in terms of strain effects \citep{Fuchs2008}\@. In order to directly identify such effects
and, thus, to shed light on the mechanism leading to these intriguing magnetic properties, near edge x-ray absorption
fine structure (NEXAFS) measurements have been performed at the Co $L{}_{2,3}$ and the O \emph{K} edges. These
measurements benefit, on one hand, from the sensitivity of the Co $L{}_{2,3}$ edge on the valence and the spin state of
the Co ion, and, on the other hand, from the fact that the unoccupied density of states (DOS) at the O \emph{K} edge
contains information about precisely those orbitals involved in the chemical bonding. Our results show that the spin
state is frozen in the E-LCO films while in the P-LCO films there are spectral changes similar to those seen in bulk
LCO\@.

\section{Experimental}

E-LCO and P-LCO films were grown on <001> oriented LSAT substrates with a film thickness of about 100~nm by pulsed laser
deposition using stoichiometric sinter targets of the corresponding compound. To ensure clean sample surfaces the
samples were grown \emph{in-situ} prior to measurements. The growth conditions were the same as those reported in Ref.\
\citep{Fuchs2007}\@. For the preparation of the P-LCO films the LSAT substrates were covered with a 20~nm thick
polycrystalline CeO$_{2}$ inhibit layer followed by the growth of the LCO film with the same parameters as for the E-LCO
films. The samples were characterized by x-ray diffraction and Rutherford backscattering spectrometry; details can be
found in Refs.\ \citep{Fuchs2007,Fuchs2008}\@. The results illustrate the high quality of the films. The NEXAFS
measurements were performed at the Institut für Festkörperphysik beamline WERA at the ANKA synchrotron light source
(Karls\-ruhe, Germany)\@. The energy resolution was set to 0.3~eV for the Co $L_{2,3}$ edge and to 0.2~eV for the O
\emph{K} edge. The Co $L{}_{2,3}$ spectra were collected in the total electron yield (TEY) mode with a probing depth of
$\approx$50 {\AA}, while the O $K$ edge spectra were recorded in the fluorescence yield (FY) mode which probes a larger
fraction, roughly 700 {\AA}, of the film thickness. Self-absorption and saturation effects in FY mode were corrected.
Some Co $L{}_{2,3}$ spectra were also taken in Auger electron yield (AEY) mode, which is considerably more
surface-sensitive than TEY and probes, in this case, about 14 {\AA} deep
\citep{Tanuma1994,Seah1998}\@. Comparison with TEY spectra should clearly bring out
contributions, if present, from possible surface effects like reconstruction, contamination, or off-stoichiometry.
However, the spectra measured with TEY and AEY were always identical, indicating that surface effects do not play a
significant role in these compounds when prepared and examined under these conditions. Photon energy calibration was
ensured by simultaneously measuring the O \emph{K} and the Ni $L{}_{3}$ peak positions of a NiO single crystal and
comparing them with literature \citep{Reinert1995}.

\section{Results and discussion}

\begin{figure}[t]
\includegraphics[width=0.475\textwidth]{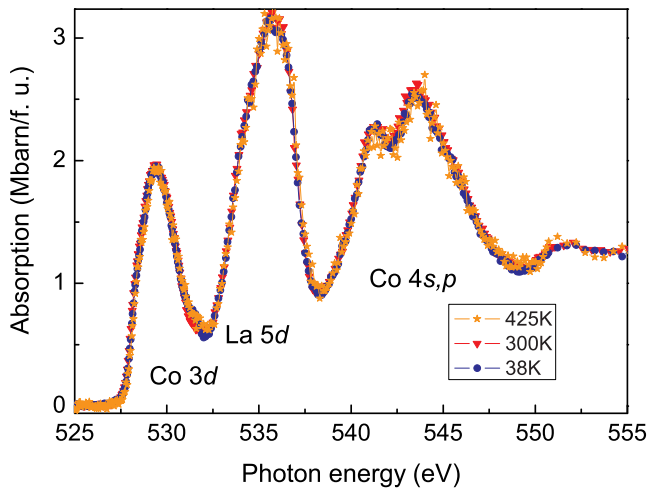}

\caption{\label{fig:EpiOK}(color online) O 1s x-ray absorption spectra of an epitaxial thin film of LaCoO$_{3}$ taken at
different temperatures. Over the whole recorded temperature range there is no change in the spectral weight
distribution.}

\end{figure}
In NEXAFS, electrons from core levels are excited into unoccupied states above the Fermi level. In the case of the O
\emph{K} edge transitions of O 1$s$ electrons to the unoccupied O 2$p$ states hybridized with Co 3$d$, La 5$d$, and Co
4$s$,$p$ states are studied. Specific peaks and features of the O \emph{K} edge spectrum can be assigned to these hybrid
states \citep{Abbate1993}, {\em cf}.\ also Fig.\ \ref{fig:EpiOK}\@. An interpretation of these absorption spectra as a
partial unoccupied density of states (DOS) is possible, mainly because the screening is good and minimizes effects of
the 1$s$ core hole \citep{Groot1989}\@. If, on the other hand, the overlap of the core hole with the final states is
significant and the screening is weak, as is the case for the 2$p$ core hole in the Co $L_{2,3}$ edge transition, the
shape of the spectra is reduced to a single spin-orbit split ``white line'' and is governed by atomic-like multiplet
effects. For their interpretation crystal field and charge transfer effects have to be taken into account
\citep{Groot2005}.

\subsection{O K edge}

\begin{figure}[b]
\includegraphics[width=0.475\textwidth]{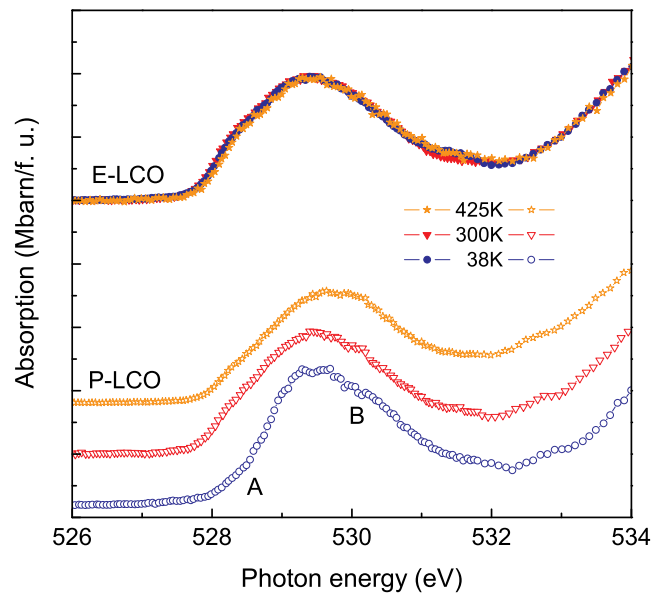}

\caption{\label{fig:OKdetail}(color online) The O $K$ low-energy region for the E-LCO and the P-LCO thin films. Open
symbols: P-LCO film. (The traces for different temperatures are offset for clarity.) Full symbols: E-LCO film strained
by the substrate. }

\end{figure}

Fig.\ \ref{fig:EpiOK} shows the results of temperature-dependent O~1\emph{s} NEXAFS for the E-LCO film. Over the whole
energy range and for all applied temperatures the spectral shape clearly does not change for this sample. We will now
focus on the energy region around 529~eV (``pre-peak'' region) since it reflects the O~2\emph{p} orbitals hybridized
with Co~3\emph{d} orbitals of $t_{2g}$ and $e_{g}$ character. Their centers of gravity are offset in energy by slightly
less than 1 eV, consistent with Ref.\ \citep{Saitoh1997}\@. This region is shown in more detail in Fig.\
\ref{fig:OKdetail}.
The pre-peak intensity is, hence, essentially proportional to the number of holes in $t_{2g}$- and $e_{g}$-derived
states.

The spectra of the P-LCO film exhibit clear evidence for a (partial) rearrangement of the electrons with temperature
between the different O$2p$-Co$3d$ states: For 38~K, the $t_{2g}$-derived spectral weight (Region A in Fig.\
\ref{fig:OKdetail}) is very low, indicating a high electron filling of the corresponding levels. At room temperature and
above \endnote{We note that the total spectral weight is somewhat reduced for the 425~K measurement. This reflects the
fact that due to a temperature-induced stretching of the Co-O bond lengths \citep{Radaelli2002}, the overlap of the O
2\emph{p} with the Co 3\emph{d} levels is reduced.}, a significant amount of spectral weight has shifted from the
``$e_{g}$ area'' (Region B) to the ``$t_{2g}$ area'' (Region A)\@. This type of rearrangement is clearly consistent with
a contribution of higher spin states that increases with temperature at the expense of the LS state predominant at low
temperatures. In other words, the Co ions are undergoing a spin-state transition similar to the situation for bulk
LCO\@. We find it interesting to note that a recent first-principles calculation \citep{Klie2007} predicts such a
rearrangement with energy shift only if the higher spin state is HS; if it is IS no energy shift compared to LS seems to
be involved.

The behavior of the epitaxial film, E-LCO, is entirely different. The spectra for the whole temperature range are almost
point-per-point equivalent; the spectral shape even for T~$\approx$~40~K resembles closely the spectrum for the
polycrystalline film, P-LCO, at room temperature and above. In particular, this means that the strain imposed on the
epitaxial film by the substrate lattice mismatch ``fixes'' its spin state very rigidly, preventing it from assuming the
pure LS state known from bulk LCO below 35~K\@ and preserving a constant spin-state configuration containing higher spin
states (along with LS) for all temperatures studied.
\begin{figure}[t]
\includegraphics[width=0.475\textwidth]{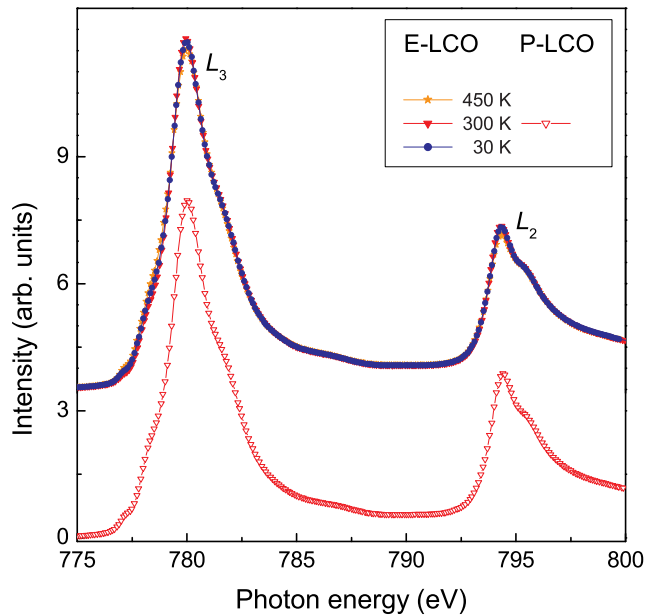}

\caption{\label{fig:EpiL3}(color online) Upper panel: Co $L_{2,3}$ NEXAFS taken at different temperatures for E-LCO\@.
No discernible spectral changes occur with temperature. Lower panel: Co L$_{2,3}$ NEXAFS for P-LCO at 300 K\@. This
trace is remarkably similar to the E-LCO spectra of the upper panel. (For clarity, the P-LCO 300 K data in the lower panel are offset 
from the temperature-dependent E-LCO spectra in the upper panel.)}

\end{figure}

\subsection{Co \emph{L} edge}

\begin{figure}[t]
\includegraphics[width=0.475\textwidth]{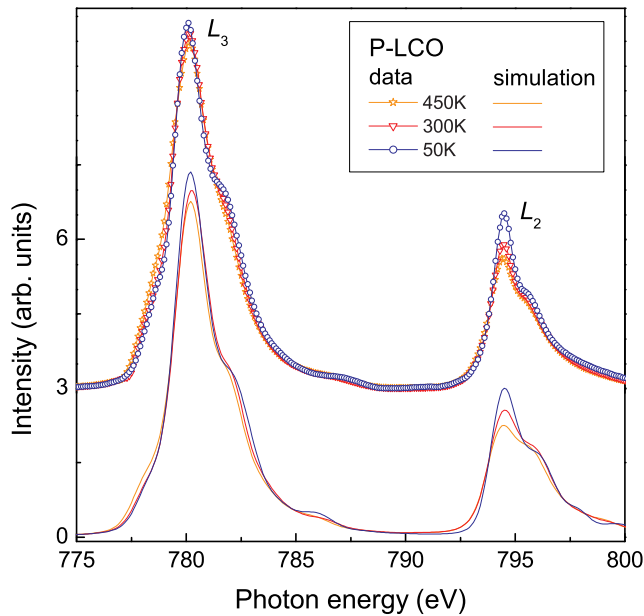}

\caption{\label{fig:polyCoL}(color online) Co $L_{2,3}$ NEXAFS taken on P-LCO films (upper panel)\@. Spectral background
and a small spectral Co$^{2+}$ contribution centered around 777~eV ({\em cf}.\ lower panel in Fig.\ \ref{fig:EpiL3})
were subtracted. The spectrum taken at 50~K exhibits two important changes over the other spectra: a pronounced shoulder
at about 781.5 eV and a significantly higher ratio between maximum and shoulder at the $L_{2}$ edge. Lower panel: Atomic
multiplet simulations using a mixture of LS and HS spectra. The spectra of the upper panel are well described by the
choice of parameters.}

\end{figure}

The temperature-dependent Co $L_{2,3}$ spectra of the LaCoO$_{3}$ samples are shown in Fig.\ \ref{fig:EpiL3} for the
E-LCO film and in Fig.\ \ref{fig:polyCoL} for the P-LCO film. The spectra correspond in first order to transitions of
the type Co 2\emph{$p^{6}$}3$d^{n}$ $\rightarrow$ Co 2$p^{5}$3$d^{n+1}$ ($n$=6 for Co$^{3+}$)\@. They consist of two
manifolds of multiplets situated around 780 ($L_{3}$) and 795 eV ($L_{2}$) and separated by the spin-orbit splitting of
the Co 2\emph{p} level. For the P-LCO films of Fig.\ \ref{fig:polyCoL}, a small spectral Co$^{2+}$ contribution centered
around 777~eV ({\em cf}.\ lower panel in Fig.\ \ref{fig:EpiL3}) was subtracted. Although this Co$^{2+}$ contribution had
been minimized by optimizing growth conditions it could not be completely avoided and might be due to some residual
oxygen deficiency.

For the P-LCO films, spectral changes with temperature are obvious: While for 50~K, the spectrum exhibits a pronounced
shoulder at about 781.5 eV and a high ratio between maximum and shoulder at the $L_{2}$ edge, both features are
considerably reduced for the spectra at higher temperatures. It is known that for bulk LaCoO$_{3}$ \citep{Abbate1993},
the shape of the Co $L_{2,3}$ edge changes with temperature, reflecting the spin-state transition occurring in bulk
material. The P-LCO spectra of Fig.\ \ref{fig:polyCoL} are similar to the earlier results of Ref.\
\citep{Abbate1993}, and the changes observed are, thus, indicative of an analogous spin-state transition from a
predominantly LS state at 50~K to a state with increasing admixture of higher spin states at 300 and 450~K\@. P-LCO is
semiconducting, and it turned out impossible to measure this sample at temperatures below 50~K due to strong charging
effects. The conductivity might further be reduced for low temperatures by the increasing fraction of Co ions in the LS
state.

\begin{figure}[h]
\includegraphics[width=0.475\textwidth]{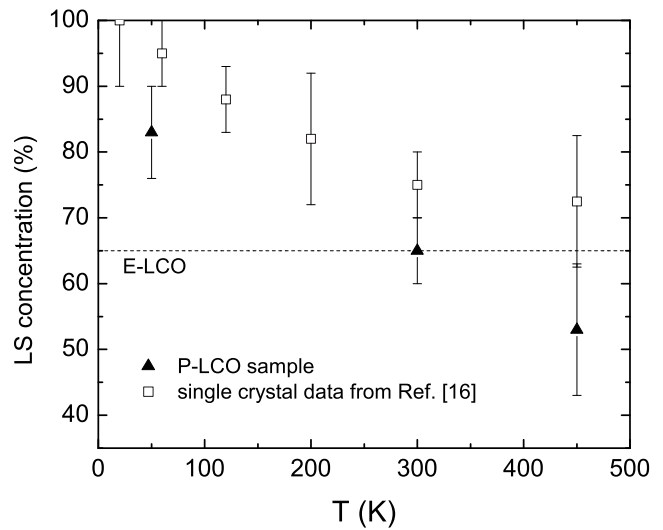}

\caption{\label{fig:HScon}Fraction of Co$^{3+}$ LS as extracted from the multiplet simulations of the NEXAFS spectra for
P-LCO, compared with results from Ref.\ \citep{Haverkort2006} for single-crystalline LCO\@. The E-LCO spectra are very
similar to the P-LCO result at 300 K, and the corresponding LS fraction is included as a dashed line.}

\end{figure}
In order to obtain a more quantitative idea of the temperature-dependent spin-state fractions the P-LCO spectra were
modeled by atomic multiplet calculations. The code developed by B. T. Thole \citep{Thole1987,Thole1988,Laan1988} and
maintained and further developed by F. M. F. de Groot \citep{Groot1993,Groot2005} was used to calculate spectra for
Co$^{3+}$ in $O_{\rm{h}}$ symmetry for different values of the crystal field splitting $\Delta_{CF}$ and of the charge
transfer energy $\Delta_{c}$\@. Charge transfer effects were included by admixing transitions of the type
2\emph{$p^{6}$}3$d^{7}\underline{L}$$\rightarrow$ 2$p^{5}$3$d^{8}\underline{L}$, where $\underline{L}$ denotes a hole in the oxygen ligand. Remarkable seems the fact that the ground state has IS character only if the spin-orbit coupling of the 3\emph{d} levels is switched off in the calculations, which in turn means that the IS lies about 0.1~eV in energy
higher than the LS and HS levels at their crossing point (at $\Delta_{CF} \approx 1.7$~eV)\@. This implies that it is
also not possible to populate the IS state by temperature, at least not in the range reached by our experiments.
Although the present calculations and those of Ref.\ \citep{Haverkort2006} are somewhat different and differ in
resulting details, this is an aspect where both agree. Calculations similar to ours have been performed in Ref.\
\citep{Groot1990}, and the authors find the HS/LS crossing point at $\Delta_{CF} \approx 2.1$~eV\@. Our energy scale is
shifted to lower values compared to those calculations by the inclusion of charge-transfer effects \citep{Groot2005}\@.

The temperature-dependent spectra of the P-LCO sample were modeled using different ratios of Co$^{3+}$ HS and LS
spectra. For the reason just pointed out the IS state does not contribute in the present simulations. Nevertheless, we
do not have direct and unambiguous experimental evidence for or against IS or HS in the spectra and, thus, will continue
to address non-LS states inclusively as ``higher'' spin states. Fig.\ \ref{fig:polyCoL} shows the fit results (lines)
compared to the data (symbols)\@. The simulations describe the measured spectra very nicely in many details, especially
regarding the peak-to-shoulder ratio at the $L_{2}$ edge which appears to be very sensitive to the amount of higher spin
states in the sample.

The concentration of the LS Co$^{3+}$ extracted from our fits is shown in Fig.\ \ref{fig:HScon} (triangles) and compared
to the concentrations extracted in Ref.\ \citep{Haverkort2006} from single-crystal data. Our LS values are lower by
about 10~\% than the results obtained for the single crystal, which might reflect grain-boundary or residual strain
effects of the polycrystalline material. However, we find the same temperature-dependent development of the LS
concentration.

For E-LCO, the situation is again very different from P-LCO: Only very little changes with temperature are detectable in
Fig.\ \ref{fig:EpiL3}, like the absolute amplitude for both the $L_{3}$ and $L_{2}$ peaks. These small changes can be
explained by an increased phonon broadening at higher temperatures. The shape of the spectra taken at all temperatures
is very similar to that of P-LCO at 300 K $-$ see also the comparison between upper and lower panels in Fig.\
\ref{fig:EpiL3}\@. The constant spectral shape means, in particular, that no spin-state changes are visible. This agrees
perfectly with the observations made at the oxygen edge. It also shows, {\em cf}.\ Fig.\ \ref{fig:HScon}, that for E-LCO
at low temperature the amount of Co$^{3+}$ in the LS state is significantly lower than at similar temperatures in
P-LCO\@. A low LS population and, conversely, a significant population in higher spin states is clearly a necessary
prerequisite for ferromagnetism.

It remains to be clarified why higher spin states are more abundant in E-LCO at low temperature. We observe that, in
contrast to the largely relaxed structural state of P-LCO, the LSAT substrate imposes strain on E-LCO, leading, among
other things, to an increased unit-cell volume (56 {\AA}$^3$ versus 54 {\AA}$^3$ for E-LCO and P-LCO, resp.)
\citep{Fuchs2007}\@ and to increased $a$,$b$ lattice parameters (3.87 {\AA} vs.\ 3.80 {\AA})\@. Geometry indicates that
this will result in a reduced tilt of the CoO$_{6}$ octahedra in LaCoO$_3$ \citep{Fuchs2008}\@. This, in turn, enhances
the overlap of the $e_g$-symmetry 3$d$ orbitals with the O 2$p$ orbitals, whereas the hybridization of the $t_{2g}$
states with the O 2$p$ orbitals is reduced. Density functional calculations \citep{Kn'ivzek2005} demonstrate a resulting
stabilization of the higher spin states; this is also consistent with the larger ionic radius of the Co$^{3+}$ ions in
IS and HS \citep{Radaelli2002}\@.

\vfill

\section{Summary and Conclusions}

Both O $K$ and Co $L_{2,3}$ NEXAFS results clearly demonstrate that for E-LCO thin films grown on LSAT, a significant
and temperature-independent fraction of higher spin states is present. This is a necessary prerequisite for
ferromagnetism and, thus, helps to explain the magnetic behavior of E-LCO\@. In contrast, P-LCO exhibits
temperature-dependent spin-state transitions in analogy to bulk LCO, resulting in a very high LS population at low
temperature and, thus, paramagnetic behavior.

\begin{acknowledgments}
Experimental help by B. Scheerer is gratefully acknowledged. We acknowledge the ANKA Ångströmquelle Karlsruhe for the
provision of beamtime. Part of this work was supported by the German Science Foundation (DFG) in the framework of the DFG Research Unit 960 ``Quantum Phase Transitions''.
\end{acknowledgments}

\bibliographystyle{apsrev}
\bibliography{PRB}

\end{document}